\documentclass[aps,prl,twocolumn,showpacs,amsmath,amssymb,superscriptaddress]{revtex4}
\usepackage{graphicx}
\usepackage{amssymb}
\usepackage{natbib}
\usepackage{color}

\newcommand{\qq}{\mathbf{q}}

\newcommand{\MM}{\mathbf{M}}

\newcommand{\beq}{\begin{eqnarray}}
\newcommand{\eeq}{\end{eqnarray}}

\begin{document}

\title{Avoided quantum criticality and magnetoelastic coupling in BaFe$_{2-x}$Ni$_{x}$As$_{2}$}
\author{Xingye Lu}
\affiliation{Beijing National Laboratory for Condensed Matter
Physics, Institute of Physics, Chinese Academy of Sciences, Beijing
100190, China}
\affiliation{ Department of Physics and Astronomy,
The University of Tennessee, Knoxville, Tennessee 37996-1200, USA }
\author{H. Gretarsson}
\affiliation{Department of Physics, University of Toronto, 60 Saint George Street, Toronto, Ontario M5S 1A7, Canada}
\author{Rui Zhang}
\affiliation{Beijing National Laboratory for Condensed Matter
Physics, Institute of Physics, Chinese Academy of Sciences, Beijing
100190, China}
\author{Xuerong Liu}
\affiliation{Condensed Matter Physics and Materials Science Department,
Brookhaven National Laboratory, Upton, New York 11973, USA
}
\author{Huiqian Luo}
\affiliation{Beijing National Laboratory for Condensed Matter
Physics, Institute of Physics, Chinese Academy of Sciences, Beijing
100190, China}
\author{Wei Tian}
\affiliation{Quantum Condensed Matter Division, Oak Ridge National Laboratory, Oak Ridge, Tennessee 37831, USA
}
\author{Mark Laver}
\affiliation{Laboratory for Neutron Scattering, Paul Scherrer Institute, CH-5232 Villigen, Switzerland
}
\affiliation{
Department of Physics, Technical University of Denmark, DK-2800 Kongens Lyngby, Denmark
}
\author{Z. Yamani}
\affiliation{Canadian Neutron Beam Centre, National Research Council, Chalk River, Ontario, K0J 1P0 Canada
}
\author{Young-June Kim}
\affiliation{Department of Physics, University of Toronto, 60 Saint George Street, Toronto, Ontario M5S 1A7, Canada}
\author{A. H. Nevidomskyy}
\affiliation{Department of Physics and Astronomy, Rice University, Houston, Texas 77005, USA
}
\author{Qimiao Si}
\affiliation{Department of Physics and Astronomy, Rice University, Houston, Texas 77005, USA
}
\author{Pengcheng Dai}
\email{pdai@utk.edu}
\affiliation{ Department of Physics and Astronomy,
The University of Tennessee, Knoxville, Tennessee 37996-1200, USA }
\affiliation{Beijing National Laboratory for
Condensed Matter Physics, Institute of Physics, Chinese Academy of
Sciences, Beijing 100190, China}

\date{\today}

\begin{abstract}
We study the structural and magnetic orders
in electron-doped
BaFe$_{2-x}$Ni$_{x}$As$_2$ by
high-resolution synchrotron X-ray and neutron scatterings.
Upon Ni-doping $x$,
the nearly simultaneous tetragonal-to-orthorhombic structural ($T_s$)
and antiferromagnetic ($T_N$)  phase transitions in BaFe$_2$As$_2$
are gradually suppressed and separated, resulting in $T_s>T_N$  with increasing $x$
as was previously observed.
However,
the temperature separation between
$T_s$ and $T_N$
decreases with increasing $x$
for $x\geq 0.065$, tending towards a quantum bi-critical point near optimal superconductivity at $x\approx 0.1$.
The
zero-temperature transition is preempted
 by the formation of a secondary incommensurate magnetic phase in the region $0.088\lesssim x \lesssim 0.104$,
 resulting in a finite value of $T_N  \approx T_c+10$~K
above the superconducting dome
around
 $x\approx 0.1$.
Our results imply an
 avoided quantum critical point,
which is expected to strongly influence the properties of both
 the normal and superconducting states.
\end{abstract}

\maketitle

A determination of the structural and magnetic phase diagram in correlated electron materials is important
for understanding their underlying electronic excitations.
In the iron pnictides, superconductivity arises at the border of both antiferromagnetic (AF) and structural orders
\cite{kamihara,cruz,qhuang,mgkim11,dai}.
This motivates the exploration of quantum critical points, where the transition temepratures for such orders
are continuously suppressed to zero by a non-thermal control parameter.
For the iron pnictide superconductors derived from electron or hole doping of
their parent compounds,
the most heavily studied materials are probably the electron-doped BaFe$_{2-x}T_x$As$_2$ (where $T=$ Co, Ni)
because of the availability of high-quality single crystals \cite{sefat,ljli,nni08,jhchu09,prozorov,clester,dkpratt,christianson09,mywang11,dkpratt11,hqluo,mgkim12,nandi}.
In the undoped state, BaFe$_2$As$_2$ exhibits a tetragonal-to-orthorhombic structural transition at temperature
$T_s$ and an AF phase transitions below nearly the same temperature $T_N\approx T_s\approx 138$ K \cite{qhuang,mgkim11}.
Upon electron-doping of  BaFe$_2$As$_2$ via partially replacing Fe by Co or Ni, various experiments, including
transport \cite{nni08,jhchu09}, neutron \cite{clester,dkpratt,christianson09,mywang11,dkpratt11,hqluo},
and high-resolution X-ray scattering \cite{mgkim11,nandi} reveal
that the structural ($T_s$) and magnetic ($T_N$) phase transition
temperatures in BaFe$_{2-x}T_x$As$_2$ gradually decrease and separate
with increasing $x$, and have $T_s>T_N$ for all doping levels.  In the initial X-ray \cite{prozorov}
and neutron \cite{clester} scattering work on BaFe$_{2-x}$Co$_x$As$_2$, it was suggested
that the separated $T_s$ and $T_N$ smoothly extend into the superconducting dome, resulting in distinct
structural and magnetic quantum critical points at different $x$. Subsequent X-ray \cite{nandi} and neutron \cite{dkpratt,christianson09,mywang11} scattering
experiments on superconducting BaFe$_{2-x}T_x$As$_2$ samples with coexisting AF order revealed that
superconductivity actually competes with the static
 AF order and lattice orthorhombicity.  As a consequence, the smoothly decreasing $T_s$ and $T_N$ are
 reported to bend back below $T_c$, and
the orthorhombic structure above $T_c$ for optimally doped sample evolves back
to a tetragonal structure  well below $T_c$ (termed the ``re-entrant'' tetragonal phase) \cite{nandi}.

\begin{figure}[t]
\includegraphics[scale=.35]{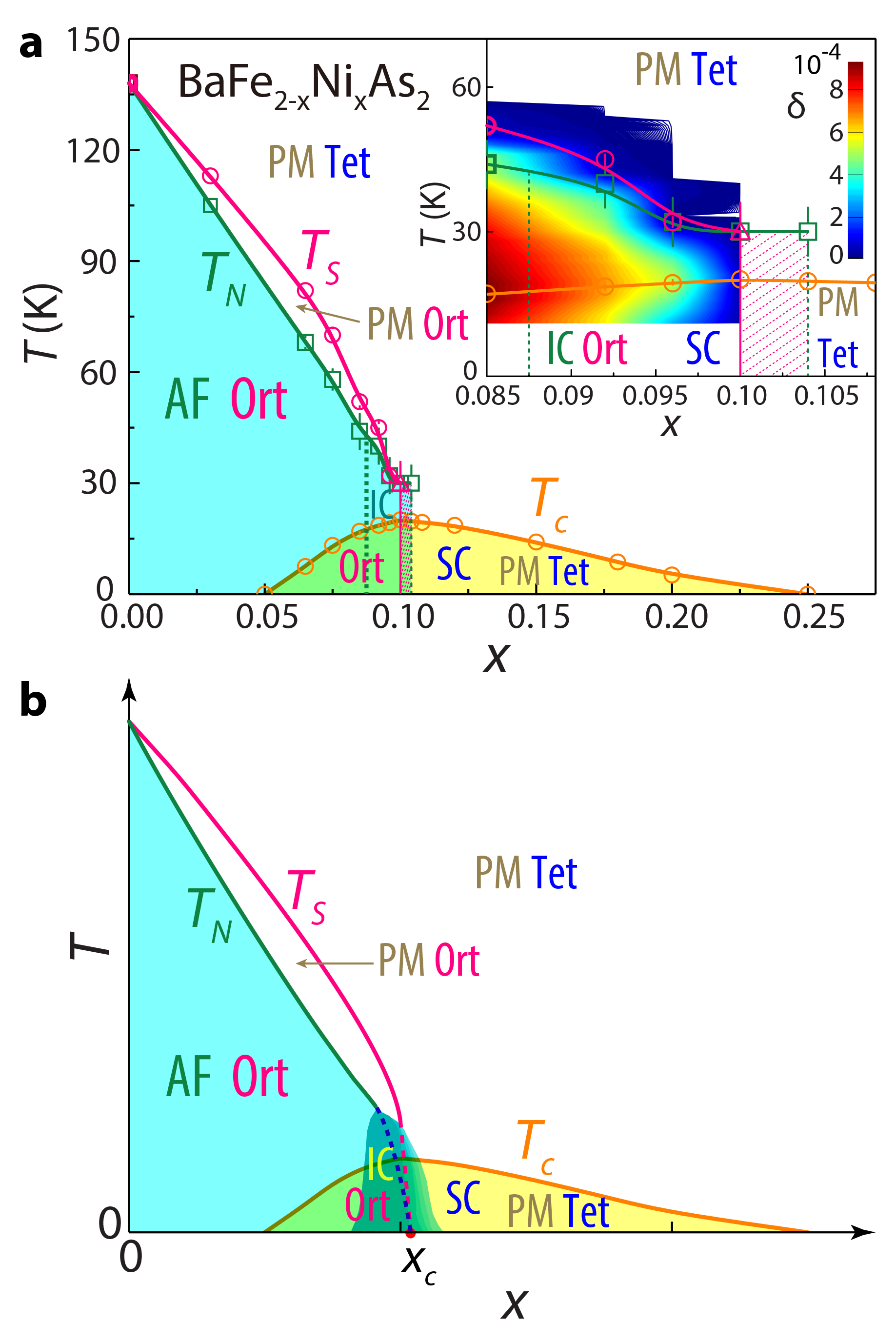}
\caption{
(a)Electronic phase diagram of
BaFe$_{2-x}$Ni$_{x}$As$_{2}$ as a function of Ni doping $x$ as determined from our neutron
and X-ray scattering
experiments. The PM Tet, PM Ort, AF Ort, IC Ort are
paramagnetic tetragonal, paramagnetic orthorhombic, commensurate AF orthorhombic, incommensurate
AF orthorhombic
phases, respectively. The AF Ort, IC Ort, and PM Tet structures in the superconducting (SC) phase
are clearly marked.  The inset shows the expanded view of $T_s$, $T_N$,
$T_c$, and temperature dependence of the orthorhombic lattice distortion order parameter
$\delta=(a_o-b_o)/(a_o+b_o)$.
The dashed region in the inset indicates the presence of a single Gaussian structural peak.
(b) Schematic theoretical phase diagram for an avoided quantum bi-critical point.
}
\end{figure}

  Although previous neutron \cite{clester,dkpratt,christianson09} and X-ray diffraction
\cite{nandi} experiments have established the magnetic
and structural phase transitions in BaFe$_{2-x}$Co$_x$As$_2$, similar measurements have not been carried out
on BaFe$_{2-x}$Ni$_x$As$_2$. In this Letter, we describe neutron and X-ray scattering studies of
structural and magnetic phase transitions in
BaFe$_{2-x}$Ni$_x$As$_2$, focusing on materials near
optimal superconductivity [Fig. 1(a)].  While neutron scattering experiments on BaFe$_{2-x}T_x$As$_2$ revealed
a commensurate-to-incommensurate AF phase transition near optimal superconductivity \cite{dkpratt11,hqluo,mgkim12},
much remains unknown about the temperature and doping evolution of
the orthorhombic lattice distortion for samples with an incommensurate
AF order.  Here, we find that $T_s>T_N$ for samples with commensurate AF order ($x\leq 0.065$),
similar to the earlier results on BaFe$_{2-x}$Co$_x$As$_2$ \cite{clester,dkpratt,christianson09,nandi}.
However, $T_s$ and $T_N$ tend to re-converge for larger values of $x$: $T_s-T_N$ decreases
for $x>0.065$.  This implicates a quantum bi-critical point
at $T=0$, which is interrupted by a secondary short-range incommensurate AF order
with very small ordered moment \cite{hqluo}.
The resulting overall phase diagram is illustrated schematically in Fig. 1(b).
Our results are important to clarifying the nature of the purported quantum critical point
in the carrier-doped iron pnictides, as inferred from the NMR~\cite{Ning,Ning10}, thermoelectric~\cite{Gooch09},
and ultrasonic~\cite{yoshizawa}  measurements, as well as its connection with the quantum
critical point of the iso-electronically tuned iron pnictides that was predicted
by theory ~\cite{Si-PNAS}
and observed by extensive experiments~\cite{delaCruz10,kasahara}.

\begin{figure}[t]
\includegraphics[scale=.5]{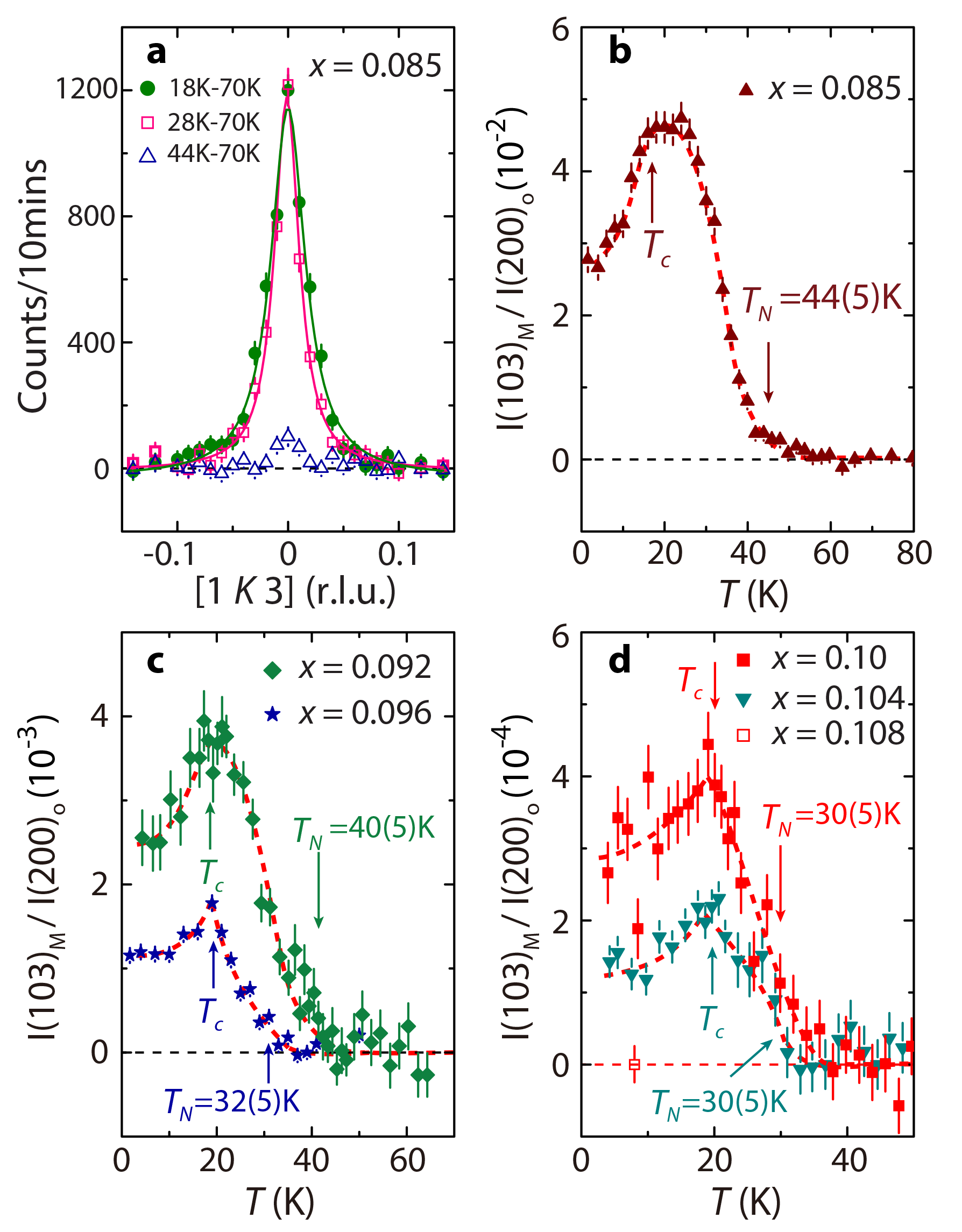}
\caption{(a)
Transverse scans along the $[1,K,3]$ direction at different temperatures
for BaFe$_{2-x}$Ni$_{x}$As$_2$ with $x=0.085$.  The magnetic scattering
each temperature was obtained by subtracting the
$T=70$ K data as background.
The change of the peak width
between 18 K and 28 K indicates the emergence of the short range incommensurate
AF order. Temperature dependence of the AF $(1,0,3)$ peak normalized to
the weak $(2,0,0)_o$ nuclear Bragg peak intensity for (b) $x=0.085$,
(c) $x=0.092,0.096$, and (d) $x=0.1,0.104$ and 0.108.
The $T_N$'s and $T_c$'s are marked by vertical
arrows.  Although there are two-orders of magnitude
magnetic scattering
intensity reduction from $x=0.085$ to 0.0104,
the $T_N$'s of the materials only decrease from $T_N=44\pm 5$ K
to $30\pm 5$ K.  The data at 7 K for $x=0.108$ was obtained by subtracting 50 K data
as background.
 }
 \end{figure}

We have carried out neutron scattering experiments on BaFe$_{2-x}$Ni$_{x}$As$_2$
with $x=0.085,0.092,0.096,0.1,0.104$ and 0.108 using
RITA-II cold neutron triple-axis spectrometer in Paul-Scherrer Institute, HB-1A thermal triple axis spectrometer
at High-Flux Isotope Reactor (HFIR), Oak Ridge National Laboratory, and C5 triple-axis spectrometer at the
Canadian Neutron
Beam Centre, Chalk River Laboratories \cite{hqluoprb}.
We have also performed high-resolution
synchrotron X-ray diffraction (XRD) experiments on identical BaFe$_{2-x}$Ni$_{x}$As$_2$ samples
using beam line X22C at
the National Synchrotron Light Source (NSLS), Brookhaven National Laboratory.
The details of the experimental procedure are given in the Supplementary Material \cite{note}.
Although neutron scattering probes the bulk sample whereas
the length scale for XRD is typically about $\sim$5 micron \cite{rutt},
both techniques are measuring the intrinsic properties of these materials.

\begin{figure}[t]
\includegraphics[scale=.4]{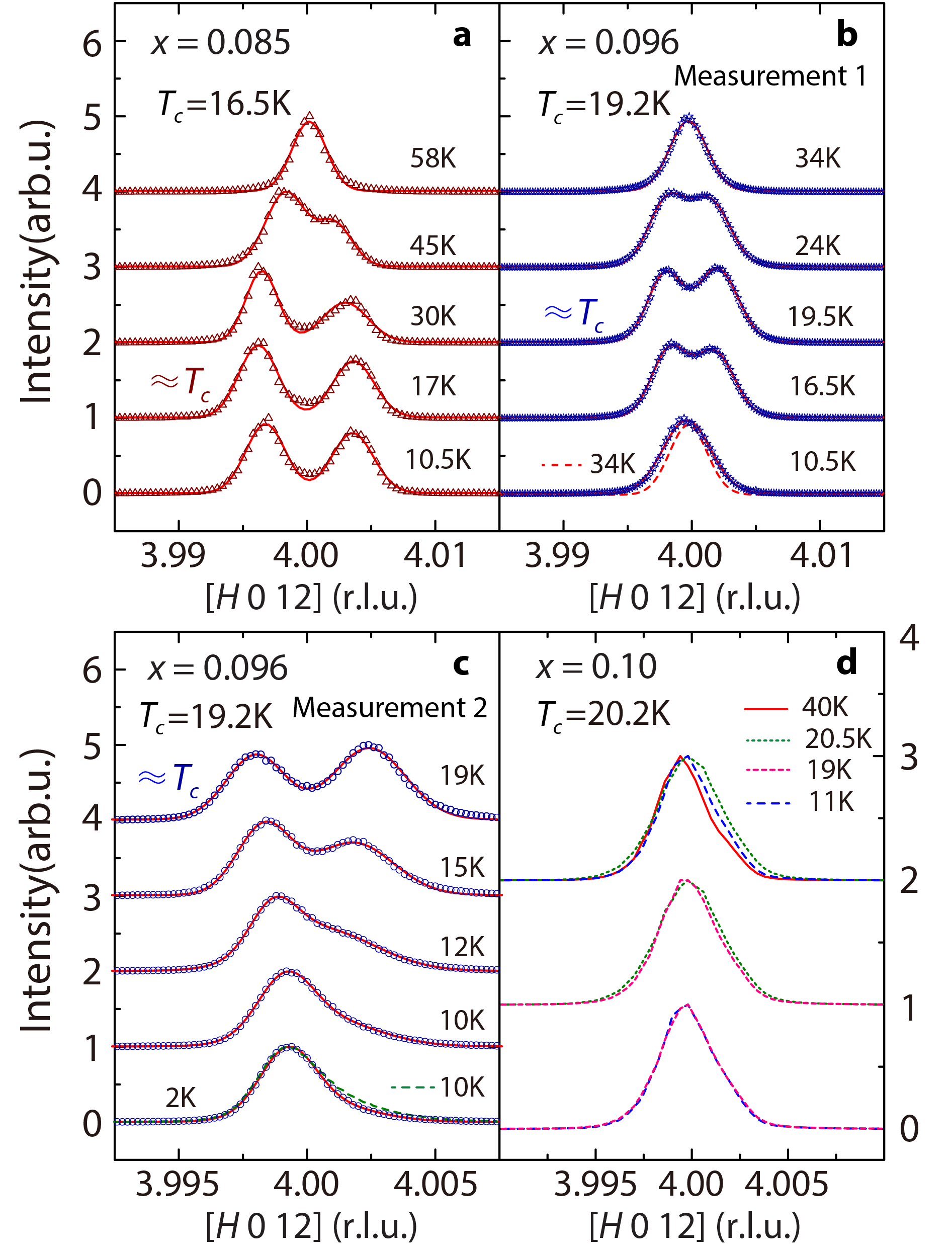}
\caption{Temperature evolution of the orthorhombic $(4, 0, 12)$ and $(0, 4, 12)$ Bragg peaks
for BaFe$_{2-x}$Ni$_x$As$_2$. Data in (a) is for $x=0.085$, (b) $x=0.096$ down to 10 K,
(c) $x=0.096$ down to 2 K, and (d) $x=0.1$ where one can only see peak broadening due to
orthorhombic lattice distortion.
These measurements were performed with $E_i=10$ keV synchrotron X-ray.
The data were collected while warming system from base temperature to a temperature well above $T_s$.
 }
\end{figure}

We first describe the determination of the N$\rm \acute{e}$el temperatures for BaFe$_{2-x}$Ni$_{x}$As$_2$
using neutron scattering.
Figure 2(a) shows transverse scans along the $[1,K,3]$ direction at different temperatures for $x=0.085$ sample.
Consistent with earlier results \cite{hqluo},
a well-defined commensurate AF order appears below 44 K.  Figure 2(b) shows temperature dependence of the magnetic
order parameter.  Again, consistent with earlier results \cite{dkpratt11,hqluo,mgkim12},
the AF order appears approximately below $T_N=44\pm 5$ K and is suppressed from the onset of $T_c$.
Figure 2(c) plots similar data for $x=0.092$ and $0.096$, showing $T_N=40\pm 5$ and $32\pm 5$ K,
respectively \cite{hqluo}.
In the previous work on optimally electron-doped BaFe$_{1.9}$Ni$_{0.1}$As$_2$ \cite{schi}, it was suggested,
based on
cold neutron data on mosaic crystals ($\sim$0.6 g) counting 1 min/point,
that there is no measurable static AF order.  Our new measurements on the
$x\!=\!0.1$ sample ($\sim$0.34 g) with much longer counting time (30 mins/piont on HB-1A)
 reveal a weak static AF order with magnetic scattering 5 times smaller than that of $x=0.096$ [Figs. 2(c) and (d)].
Similar measurements on $x=0.104$ also show the presence of a weak static AF order, which is 50\%
smaller than
that of the $x=0.1$ sample.  In spite of their small moments, the temperature dependence of the magnetic order
parameters
for both samples indicate that their N$\rm \acute{e}$el temperatures are essentially unchanged at $T_N=30\pm 5$ K
[Fig. 2(d)].
Finally, we find no evidence of static AF order for a $x=0.108$ sample ($\sim$0.5 g)
by counting 40 mins/point on C5 [Fig. 2(d)].

 \begin{figure}[t]
\includegraphics[scale=.45]{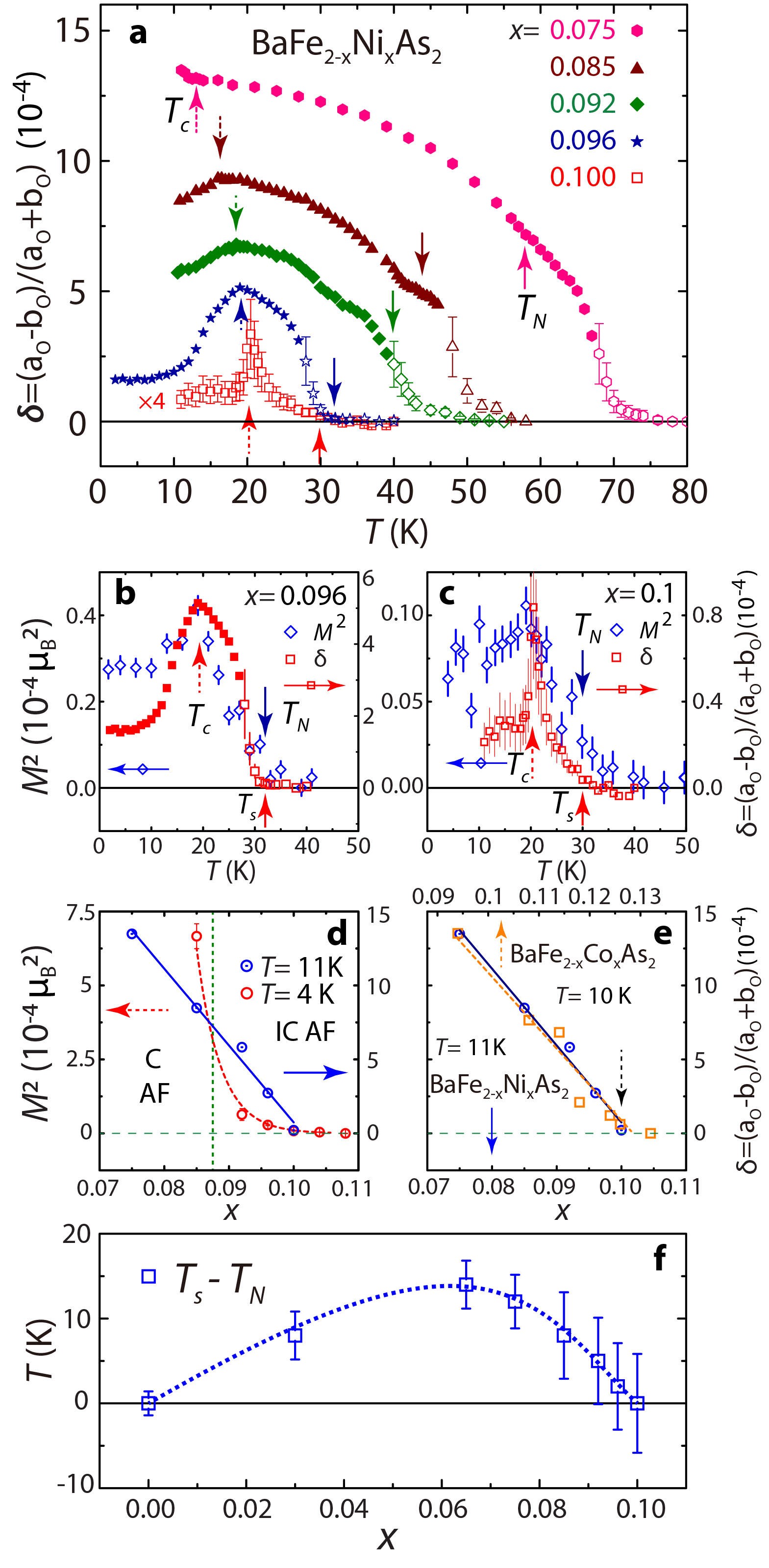}
\caption{(a) Orthorhombic lattice distortion $\delta$
as a function of temperature for BaFe$_{2-x}$Ni$_x$As$_2$.  The data denoted by the filled symbols are derived
from fitting $(4,0,12)$ and $(0,4,12)$ Bragg peaks by two peaks, while the open symbols are data obtained from
deconvolving the instrumental resolution-limited
peak at a temperature above $T_s$. The magnitude of $\delta$ for $x=0.1$
was multiplied by a factor of 4 for clarity.  The vertical arrows indicate positions of $T_N$ and $T_c$.
Comparison of the temperature dependence of the magnetic order
parameter and orthorhombic lattice distortion $\delta$ for
(b) $x=0.096$ and (c) $x=0.1$. (d) Ni-doping dependence of
the magnetic Bragg peak intensity at 11 K
and $\delta$.  The vertical dashed line indicates the
boundary between commensurate and incommensurate AF order.
(e) Comparison of Co \cite{nandi} and Ni doping dependence of $\delta$.  In
both cases we see structural quantum critical point near optimal
superconductivity at $x=0.1$.  (f) Electron doping dependence of $T_s-T_N$.
 }
\end{figure}

In order to compare the onset of orthorhombicity with antiferromagnetism,
high-resolution X-ray scattering measurements were performed on the samples identical to those used for
neutron scattering. In all cases, we carried out longitudinal scans along the $[H,0,12]$ direction.
  Figure 3(a) shows the outcome for $x=0.085$ which has a superconducting $T_c=16.5$ K.  At $T=58$ K,
  a temperature well above $T_s$, we see a single
instrumentation resolution-limited peak, consistent with a tetragonal lattice.  On cooling to $T=45$~K, 30~K and 17 K,
the single peak splits into two peaks with increasing peak separations as temperature decreases down to $T_c$.
Upon further cooling  below $T_c$, the peak separations become smaller,
as if the system turns back toward
the tetragonal structure \cite{nandi}.  Figure 3(b) shows similar temperature dependent scans for
 $x=0.096$.  Although the split peaks appear to become a single peak at $T=10.5$ K, its width is
 still larger than that in the tetragonal phase ($T=34$ K), suggesting that the nearly optimal superconductor
 has an orthorhombic lattice distortion at $T=10.5$ K.  To see how such orthorhombic lattice distortion evolves at lower temperatures,
 we carried out additional measurements using a cryostat capable of going down to 2~K.  The longitudinal $[H,0,12]$ scans
 in Fig. 3(c) show broad peaks at temperatures below 10 K, suggesting the presence of an orthorhombic lattice structure
 even at 2~K.

To quantitatively analyze the temperature dependence of the orthorhombic lattice distortion, we define
lattice orthorhombicity $\delta=(a_o-b_o)/(a_o+b_o)$, where $a_o$ and $b_o$ are lattice parameters of
the orthorhombic unit cell \cite{nandi}.  Figure 4(a) shows the temperature
dependence of $\delta$ for BaFe$_{2-x}$Ni$_{x}$As$_2$ with $x\!=\!0.075,0.085,0.092,
0.096$, and 0.1.
Figures 4(b) and 4(c) compare the ordered moment squared, $M^2$,  with the lattice orthorhombicity $\delta$,
and their similar temperature dependence suggests a strong magnetoelastic coupling.

The optimally doped $x=0.1$ sample ($T_c=20.2$ K) deserves special attention.
Its temperature dependent $[H,0,12]$ scans are shown in  Figure 3(d).
Although we can no longer see the double peaks, we observe a peak broadening that does not disappear  at low temperatures.
We therefore used the full width at half maximum (FWHM) of the peak in order to determine the lattice
orthorhombicity $\delta$, similar to the analysis of BaFe$_{2-x}$Co$_{x}$As$_2$ by Nandi~\emph{et al.} \cite{nandi}.
The deduced temperature dependence of $\delta$ is shown in Fig.~4(a) with red squares and appears
to have a sharp cusp near the superconducting $T_c$.  We conjecture that this cusp occurs
because the electron-lattice coupling results in a lattice response to the superconducting fluctuations near $T_c$.
At the lowest temperature measured, $T=11$~K, the value of $\delta$ is too small ($2\times10^{-5}$) in order to
unambiguosly claim the orthorhombicity.  However, taken together with magnetization squared for incommensurate
AF order, see Fig.~4(c), which has a similar temperature dependence, we conclude that a weak static AF order likely
coexists with orthorhombic lattice distortion in the optimally superconducting  BaFe$_{2-x}$Ni$_{x}$As$_2$,
different from the re-entrant tetragonal transition seen in BaFe$_{2-x}$Co$_{x}$As$_2$ \cite{nandi}.

Figure 4(d) shows the Ni-doping dependence of $\delta$ and the ordered moment squared $M^2$,
while Figure 4(e) compares the doping dependence of $\delta$ in BaFe$_{2-x}$Ni$_x$As$_2$ and
in the previously reported
BaFe$_{2-x}$Co$_x$As$_2$ \cite{nandi}.
The essentially continuous suppresion of both $M^2$ and $\delta$ near $x=0.1$ provides further evidence
for an extrapolated quantum critical point.
For the magnetic ordering, this represents a new understanding.  On the other hand,
for the orthorhombic distortion, the continuos suppression of $\delta$ with doping was already anticipated by
ultrasound spectroscopy measurements  \cite{fernandes,yoshizawa}.

Theoretically, this can be considered through a Landau--Ginzburg action for such a criticality,
$ S = S_M[\mathbf{M}] + S_\text{lat}[\phi] +
S_{\text{lat}-M}[\mathbf{M}, \phi]
$; the three terms, describing
the magnetic and lattice parts, respectively, and their coupling, are given in the
Supplementary Material \cite{note}.
This model resembles the previously studied  O(3)$\times$Z$_2$
model~\cite{Si-PNAS}, except that here the lattice
quantum field $\phi$ is endowed with its own dynamics and undergoes Landau damping $\Gamma_s$,
making it inherently quantum critical
with the dynamic exponent $z=3$.
In two spatial dimensions, $d+z\!=\!5$ for the $\phi$ field and $d+z\!=\!4$ for the
$\MM$ fields. Because they are above/at the upper critical dimension,
a quantum \emph{bi-critical} point for both orders is expected in the presence of
the magneto-elastic coupling $\eta$.
This is similar to the result of the O(3)$\times$Z$_2$
model~\cite{Si-PNAS}, and is indicated schematically in Fig.~1(b).
Indeed, as noted above, our measurements find that $T_N$ and $T_s$ get closer to each other
as the quantum critical point is approached [see Figs.~1(a) and 4(f)] and the two order parameters
disappear at the same point [Fig.~4(d)]. However, the appearance of an emergent incommensurate magnetism
at $x\approx 0.088$ severely  reduces the scattering rate $\gamma$ and $\Gamma_s$ (in addition
to modifying other parameters of the effective theory), thereby eliminating the quantum critical point.
A quantum critical point preempted by an emergent order
is often referred to as ``avoided'' quantum criticality~\cite{coleman,haule, harrison}.

From direct measurements of the
order parameters for both the AF and structural transitions,
our results provide a solid basis
for quantum criticality in carrier-doped iron pnictides, which has so far been indirectly deduced from
the temperature
dependences of magnetic, transport, or acoustic properties~\cite{Ning,Ning10,Gooch09,yoshizawa}.
In addition, because the primary AF order in the electron-doped iron pnictides discussed here
is commensurate,  our results suggest that
the quantum critical point arising under the carrier doping is surprisingly similar to
 that induced by iso-electronic doping ~\cite{Si-PNAS,delaCruz10,kasahara};
 the main distinction of the carrier doping is to introduce a secondary incommensurate order.
 This reveals an important universality of the underlying physics for the
 iron pnictides under carrier and iso-electronic dopings.

Summarizing the results presented in Figs.~2-4, we show in Fig.~1(a) the refined phase diagram
of BaFe$_{2-x}$Ni$_{x}$As$_2$, in agreement with the theoretically expected one  [Fig.~1(b)].
While the phase diagram is mostly consistent with the earlier work on
BaFe$_{2-x}$Co$_{x}$As$_2$ at low electron doping levels \cite{nandi},
our key new finding is that when $x$ approaches optimal doping, the magnetic and structural transition
temperatures converge to the purported quantum bi-critical point, with both order parameters
disappearing near $x\approx 0.1$ [Fig.~4(d)] as a result of magneto-elastic coupling.
However, the emergent short-range incommensurate magnetism helps the system avoid the quantum critical fate,
resulting in an apparent saturation of $T_s\sim T_N\approx 30$~K above the superconducting  $T_c$ near
optimal doping $x=0.1$, as shown in  Fig. 1(a).  These results elucidate
the quantum criticality
in the carrier-doped iron pnictides and its connection with that of the iso-electronically doped counterparts,
and reveal a rich theoretical picture that should be further explored in future work.

The work at IOP, CAS, is
supported by MOST (973 project: 2012CB821400 and
2011CBA00110) and NSFC (No.11004233). The work at UTK is supported by the
U.S. NSF-DMR-1063866. The work at Rice University is supported by
the U.S. NSF-DMR-1006985 and the Robert A.
Welch Foundation Grants Nos.~C-1411 and C-1818. Research at the University of Toronto
was supported by the NSERC and CFI. Use of the NSLS was supported by the
U.S. DOE,BES, under Contract No. DE-AC02-98CH10886. The work at
the HFIR, ORNL, was sponsored by the
Scientific User Facilities Division, BES, U.S. DOE.


\newpage
\maketitle{\textbf{Supplementary material: Avoided quantum criticality and magnetoelastic coupling in BaFe$_{2-x}$Ni$_{x}$As$_{2}$}}
\\[5mm]

Section A: {\bf Details of the neutron and X-ray scattering experiments}
\\[3mm]

Neutron scattering experiments:
The RITA-II uses a pyrolytic graphite (PG) filter before the sample
and a cold Be filter after the sample with the final neutron energy fixed at $E_f=4.6$ meV \cite{hqluo}.  HB-1A
spectrometer operates with a fixed incident neutron energy of $E_i=14.64$ meV
using a double PG monochromator. The second-order contamination in the beam was removed
by placing two PG filters located before and after the
second monochromator. A collimation of $48^\prime$-$48^\prime$-sample-$40^\prime$-$68^\prime$
from reactor to detector was used throughout the measurements.
C5 uses PG as monochromator and analyzer with fixed $E_f=14.56$ meV \cite{hqluoprb}.
The sample orientation and setup are similar to those described previously \cite{hqluo}.  Figure S1 shows the Ni-doping dependence of
the incommensurate AF order.  For BaFe$_{2-x}$Ni$_{x}$As$_{2}$ samples with $x=0.085, 0.092$, we see clear incommensurate static AF order below $T_N$.

X-ray diffraction experiments:
The monochromator was Si(111) and incident beam energy was set at
$E_i=10$ keV with spot size of $1\times1$ mm$^2$ on the samples.
In all cases, the data were collected on warming
from base temperature to a temperature well above $T_s$.

Section B: {\bf The proposed theoretical model:}
\\[3mm]
The proposed Landau--Ginzburg action is
$S= S_M + S_\text{lat} + S_{\text{lat}-M}$.
The magnetic part $S_M$ (here $M_{A/B}$
refers to  sublattice magnetization) is specified by
\begin{widetext}
\begin{eqnarray*}
S_M &= &
\int d\{\bf q\}
  \int d\{\omega\} [S_2({\bf q},\omega) + S_4 (\{{\bf q}\},\{\omega\}) + \dots], \nonumber \\
S_2 ({\bf q},\omega) &=& \sum_{\tau = A,B}\left(\alpha_m +c_M(\qq-\mathbf{Q})^2
+ \gamma|\omega|\right)\MM_\tau^2
+ (\cos q_x -\cos q_y) \MM_{A}\cdot \MM_{B}
\nonumber \\
S_4 (\{{\bf q}\},\{\omega\}) &=&  u \sum_{\tau = A,B} |\MM_\tau|^4
 + u'|\MM_A|^2\; |\MM_B|\,^2
- v |\MM_A\cdot\MM_B|^2
\end{eqnarray*}
\end{widetext}
Here, $S_2$ describes the quadratic contribution of magnetic fluctuations,
which includes a Landau damping $\gamma$
~\cite{Hertz-Millis}.
The quartic term
$S_4$ describes mode-mode interaction between magnetic fluctuations
on the same and on different magnetic sublattices;
the last term
has a negative coefficient ($-v<0$),
favoring the collinear alignment
of spins on the two sublattices
~\cite{CCL}.
There are also lattice parts:
%
\begin{eqnarray*}
 S_\text{lat} &=&  \int d\{\qq \}
  \int d\{\omega\}
 \left(\!\alpha_s + c_sq^2 + \Gamma_s\frac{|\omega|}{q}\right) |\phi|^2
 \nonumber \\
 &&+ w \int d\{\qq\}
  \int d\{\omega\}  \; \phi^4
 \nonumber \\
 S_{\text{lat}-M} &=& -\eta
  \int d\{\qq\}
  \int d\{\omega\} \; (\MM_A\cdot\MM_B)\,\phi
\end{eqnarray*}
The last term describes magneto-elastic coupling, leading to an Ising magnetic order
when the lattice orthorhombicity $\delta \equiv \langle \phi \rangle$ develops.

\newpage

\begin{figure}[t]
\includegraphics[scale=.40]{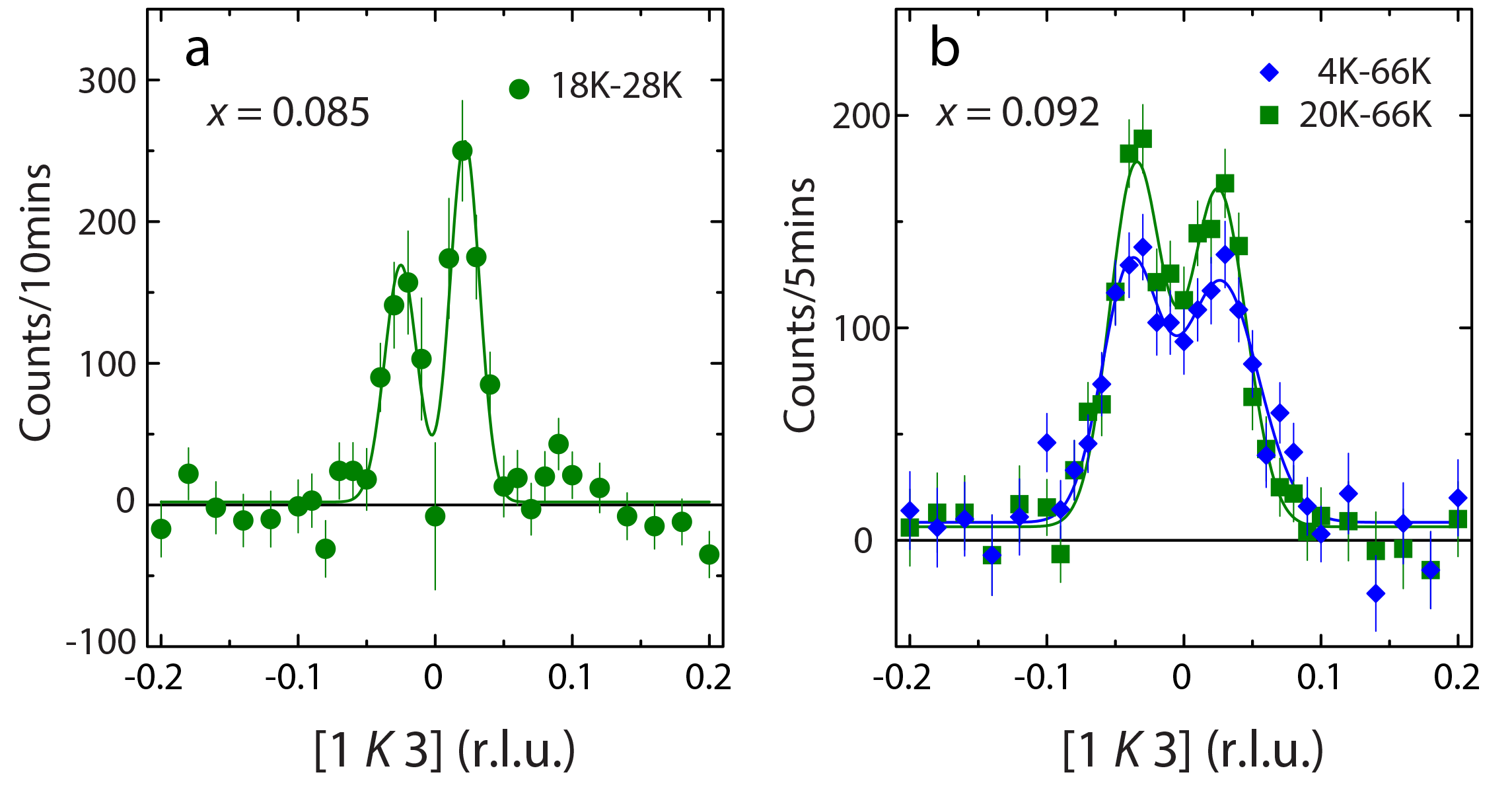}
\caption{ Ni-doping evolution of incommensurate magnetic peaks for
(a) $x=0.085$ and (b) $x=0.092$ in BaFe$_{2-x}$Ni$_{x}$As$_{2}$.  The data are obtained carrying out scans
along the $[1,K,3]$ direction at different temperatuers.
We obtain the net magnetic scattering at low temperatures by subtracting the
high-temperature ($T>T_N$) background.
}
\end{figure}

\end{document}